\documentclass[12pt,reqno,a4paper,twoside]{article}
\usepackage{amsmath,amsthm,amstext,amscd,amssymb,euscript}
\usepackage{epsf}
\renewcommand{\phi}{\varphi}

\renewcommand{\subset}{\subseteq}
\renewcommand{\emptyset}{\varnothing}

\def\1{ {\mathit{1} \!\!\>\!\! I} }

\makeatletter
\@addtoreset{equation}{section}
\makeatother

\newtheorem{ittheorem}{Theorem}
\newtheorem{itlemma}{Lemma}
\newtheorem{itproposition}{Proposition}
\newtheorem{itdefinition}{Definition}
\newtheorem{itremark}{Remark}

\newenvironment{theorem}{\addtocounter{equation}{1}
\begin{ittheorem}}{\end{ittheorem}}

\newenvironment{lemma}{\addtocounter{equation}{1}
\begin{itlemma}}{\end{itlemma}}

\newenvironment{proposition}{\addtocounter{equation}{1}
\begin{itproposition}}{\end{itproposition}}

\newenvironment{definition}{\addtocounter{equation}{1}
\begin{itdefinition}}{\end{itdefinition}}

\newenvironment{remark}{\addtocounter{equation}{1}
\begin{itremark}}{\end{itremark}}

\newcommand{\beq}{\begin{eqnarray}}
\newcommand{\eeq}{\end{eqnarray}}

\newcommand{\be}{\begin{equation}}
\newcommand{\ee}{\end{equation}}

\newcommand{\bl}{\begin{lemma}}
\newcommand{\el}{\end{lemma}}

\newcommand{\br}{\begin{remark}}
\newcommand{\er}{\end{remark}}

\newcommand{\bt}{\begin{theorem}}
\newcommand{\et}{\end{theorem}}

\newcommand{\bd}{\begin{definition}}
\newcommand{\ed}{\end{definition}}

\newcommand{\bp}{\begin{proposition}}
\newcommand{\ep}{\end{proposition}}

\newcommand{\bc}{\begin{corollary}}
\newcommand{\ec}{\end{corollary}}

\newcommand{\bpr}{\begin{proof}}
\newcommand{\epr}{\end{proof}}

\newcommand{\bi}{\begin{itemize}}
\newcommand{\ei}{\end{itemize}}

\newcommand{\ben}{\begin{enumerate}}
\newcommand{\een}{\end{enumerate}}

\newcommand{\Z}{\mathbb Z}
\newcommand{\R}{\mathbb R}

\newcommand{\C}{\mathcal C}

\newcommand{\E}{\mathbb E}

\newcommand{\pee}{\ensuremath{\mathbb{P}}}

\newcommand{\ce}{\ensuremath{\mathcal{C}}}

\newcommand{\uu}{\ensuremath{\mathcal{U}}}

\newcommand{\loc}{\ensuremath{\mathcal{L}}}
\newcommand{\iii}{\ensuremath{\mathcal{I}}}
\newcommand{\bee}{\ensuremath{\mathcal{B}}}

\newcommand{\mee}{\ensuremath{\mathcal{M}}}
\newcommand{\U}{\ensuremath{\mathcal{U}}}

\newcommand{\La}{\ensuremath{\Lambda}}
\newcommand{\la}{\ensuremath{\Lambda}}
\newcommand{\si}{\ensuremath{\sigma}}
\newcommand{\om}{\ensuremath{\omega}}

\newcommand{\QED}{\hspace*{\fill}$\Box$\medskip}

\newcommand{\tr}{\ensuremath{\mbox{Tr}}}

\newcommand{\Ga}{\ensuremath{\Gamma}}
\newcommand{\ga}{\ensuremath{\gamma}}
\newcommand{\al}{\ensuremath{\alpha}}
\newcommand{\shit}{\ensuremath{[-\|X \|, \|X \|]}}
\newcommand{\haa}{\ensuremath{A+tB}}

\def\now{
\ifnum\time<60
          12:\ifnum\time<10 0\fi\number\time am
          \else
            \ifnum\time>719\chardef\a=`p\else\chardef\a=`a\fi
          \hour=\time
          \minute=\time
          \divide\hour by 60 
          \ifnum\hour>12\advance\hour by -12\advance\minute by-720 \fi
          \number\hour:%
          \multiply\hour by 60 
          \advance\minute by -\hour
          \ifnum\minute<10 0\fi\number\minute\a m\fi}
\newcount\hour
\newcount\minute
\numberwithin{equation}{section}         

\theoremstyle{remark}

\newcommand{\caA}{{\mathcal A}}

\newcommand{\caD}{{\mathcal D}}

\newcommand{\caU}{{\mathcal U}}
\newcommand{\caV}{{\mathcal V}}

\newcommand{\un}{\underline}

\begin{document}

\title{{\bf Large deviations for quantum spin systems}}
\author{
K. Neto\v{c}n\'{y}\footnote{EURANDOM, Postbus 513,
5600 MB Eindhoven, The Netherlands, }\\
F.\ Redig\footnote{Faculteit Wiskunde en Informatica and EURANDOM, Technische Universiteit Eindhoven, Postbus 513,
5600 MB Eindhoven, The Netherlands, f.h.j.redig@tue.nl}\\
}
\maketitle
\footnotesize
\begin{quote}
{\bf Abstract:}We consider high temperature KMS states for quantum spin systems
on a lattice. We prove a large deviation principle for the
distribution of empirical averages
$\overline{X}_{\Lambda} := \frac{1}{|\Lambda|} \sum_{i\in\Lambda} X_i$,
where the $X_i$'s are copies of a self-adjoint element $X$ (level one large
deviations). From the analyticity of the generating function, we obtain the
central limit theorem.
We generalize to a level two large deviation principle for the
distribution of $\frac{1}{|\la|}\sum_{i\in\la} \delta_{X_i}$.
\end{quote}
\normalsize

{\bf Keywords:} large deviation principle, central limit theorem, boundary terms,
cluster expansion,
Goldon-Thompson inequality.

\section{Introduction}
Large deviations for classical lattice spin systems constitutes by now a rather
complete theory, see e.g.  \cite{Ellis}, \cite{olla}. In particular, for Gibbsian random fields, it is well-known that the
relative entropy density governs the large deviations of the empirical measure,
see e.g. \cite{geo}, \cite{olla}, \cite{efs}.
The relative entropy density is the Legendre transform of a generating function
which is a difference of pressures.
For instance, if one studies the large deviations of the magnetization in a Gibbs measure
with Hamiltonian $H$, one has to consider the generating function
$F(t) = P(H+ t h) -P(H)$, where $h$ is a magnetic field Hamiltonian. The Legendre
transform of $F$ gives the entropy function $I$ of the large deviations of the magnetization, i.e.,
\[
\pee_H (\frac{1}{|\la|} \sum_{i\in\la} \si_i \simeq a) = e^{-|\la| I(a)}e^{o(|\la|)}
\]
where $\si_i, i\in\Z^d$ is the value of the lattice
spin at site $i$.

For quantum lattice spin systems a similar large deviation question can be asked.
The $\si_i$ have to be replaced by self-adjoint operators $X_i$, and the probability measure $\pee_H$ has to be replaced
by a (KMS)-state. We are then interested in the ``probability"
\be\label{klak}
\omega \Bigl(1_{A} \Bigl(\frac1{|\la|} \sum_{i\in\la}X_i\Bigr)\Bigr)
\ee
where $\omega$ is a KMS state, $X_i$ are copies of an observable $X$ at
site $i$, $A$ is a Borel subset of $\shit$, and $\la\subset\Z^d$ is
a (large) volume.

Surprisingly, such probabilities
have not been considered in the literature on quantum spin systems. Laws of large numbers,
and central limit theorems have been considered, e.g. in \cite{vets}, \cite{vetsv}. It is well known that
for large volumes $\la$, the empirical average $\frac1{|\la|} \sum_{i\in\la} X_i$
is well-approximated (in the state $\omega$) by $\omega (X_0) \text{Id}$ 
provided the state $\omega$ is
mixing. Therefore, it is a natural question to ask whether the probabilities in (\ref{klak})
are of the form $\exp (-|\la| \inf_{a\in A} I(a))$, for some entropy function $I$. In the context
of non-interacting bosons, this question has been studied in \cite{Lebowitz}, later generalized in
\cite{lebspo} to weakly interacting bosons and fermions, where one considers large deviations
of the particle density.

In this paper we prove the large deviation principle for empirical averages of the form
$\frac1{|\la|} \sum_{i\in\la}X_i$ for high temperature KMS states $\omega$ (i.e., so-called
level-1 large deviations), and give a generalization to level-2 large deviations,
i.e., large deviations for distribution of the ``measures" $\loc_\la =\frac1{|\la|} \sum_{i\in\la}\delta_{X_i}$ under
the state $\omega$. The existence of the generating function of the large deviations of
$\frac1{|\la|} \sum_{i\in\la} X_i$ is not as obvious as in the classical lattice spin context
(unless $X$ commutes with the Hamiltonian of the KMS state). In fact
this generating function is not a difference of two pressures, simply obtained by perturbing
the Hamiltonian of the original KMS state by a magnetic field Hamiltonian. We show that the entropy
function obtained by proceeding as in the classical case is (strictly) larger than the true entropy
function. The reason we limit ourselves to high-temperature states is the use of a polymer expansion.
This polymer expansion can be set up because we study the large deviations of averages of
{\em one-point} observables.

Our paper is organized as follows. In section 2 we set up basic notation, specify our problem,
and state the main result of the paper. In section 3 we consider the easy case of product states,
in section 4 we compare with the classical case. In section 5 we show that the classical proof of
existence of pressure does not work if we want to show existence of the generating function for the
large deviations of empirical averages. In section 6 we set up the cluster expansion, the basic technical
tool to obtain both existence, ``boundary condition independence" and analyticity of the generating function.
In Section 7 we prove the main theorem and point out a generalization to empirical averages
of local (not necessarily single site) observables. Finally, in section 8 we prove level 2 large deviations.

\section{The Problem}
Let $M$ be a finite dimensional algebra of complex matrices. For
$\Lambda\subset\Z^d$ we define the local algebra
\begin{equation}
{\U}_\Lambda:= \otimes_{i\in\Lambda} M_i
\end{equation}
where each $M_i$ is a copy of $M$. The algebra of local observables is
defined as the inductive limit of the $\U_\Lambda$'s, and is denoted
by $\U$. Let $X\in M$ be a fixed self-adjoint element, and  consider for
$\Lambda\subset\Z^d$ the empirical average $\frac{X_\La}{|\La|}$, where
\begin{equation}
  X_\La = \sum_{i\in\Lambda} X_i
\end{equation}
and $X_i$'s are copies of $X$ in $M_i$.
Suppose
we are given a faithful state $\omega$ on $\U$. Given $A\in \U$ such that
$A= A^*$, we can consider a probability measure on the spectrum of $A$, defined
by
\begin{equation}\label{prob}
\int_{\sigma (A) } \pee_A (dx) f(x) = \omega ( f(A)),
\end{equation}
for $f:\sigma (A) \rightarrow \R$ continuous. In particular, for
$F\subset \sigma (A)$ a Borel measurable subset of the spectrum, we
have
\begin{equation}
\pee_A (F) = \omega (1_F (A) ).
\end{equation}
We call
$\pee_A$ the distribution of $A$.
Given a self-adjoint element $X\in M$, we are interested in  the probability
measures associated to the empirical averages, i.e.,
\begin{equation}
\pee^X_{\La} := \pee_{\frac{1}{|\La |} X_{\La}}.
\end{equation}
$\pee^X_{\La}$ are probability measures on $[ - \|X\| , \| X \| ]$, i.e., they
have compact support and hence always contain convergent subsequences.
If the state $\omega$ is sufficiently mixing, then $\pee^X_\Lambda$ converges weakly
to the Dirac measure $\delta_{\omega (X_0)}$, concentrating on $\omega (X_0)$.\

Therefore it is natural to ask whether the sequence
$\{ \pee^X_{\La} : \La \subset \Z^d  \}$ satisfies a large deviation principle.
This means there exists a lower-semicontinuous convex function
$I: [ - \| X \| , \| X \| ] \rightarrow \R$ such that
\begin{eqnarray}\label{LDP}
\limsup_{\La\uparrow\Z^d } \frac{1}{|\La |} \log \pee^X_{\La} (F) &\leq &
-\inf_{x\in F} I (x) \ \mbox{for} \ F\subset \R \ \mbox{closed},
\nonumber\\
\liminf_{\La\uparrow\Z^d} \frac{1}{|\La |} \log \pee^{X}_{\La} (G) &\geq &
-\inf_{x\in G} I (x) \ \mbox{for} \ G\subset \R \ \mbox{open}.
\end{eqnarray}
In some sense (\ref{LDP}) is purely a property of a particular sequence
of probability measures with compact support. Therefore, a sufficient
condition
is the existence of a differentiable generating function
\begin{eqnarray}\label{ft}
F(t)&:= & \lim_{\La\uparrow\Z^d} \log \frac{1}{|\La|}
\int \pee^X_{\La} (dx) e^{t|\Lambda| x}
\nonumber\\
&=& \lim_{\La\uparrow\Z^d} \frac{1}{|\La |} \log \omega
\left( e^{t \sum_{i\in\Lambda} X_i} \right).
\end{eqnarray}
More precisely, following \cite{Ellis}, we have the following standard
result:
\bp\label{prop1}
If for all $t\in\R$, $F(t)$ exists and is differentiable in $t$, then the large
deviation
principle
(\ref{LDP}) holds and the entropy function is
\begin{equation}
I(x) = \sup_{t\in \R} (xt- F(t)).
\end{equation}
\ep
Differentiability in $t$ can be replaced by strict convexity of $I$.
Even if $F$ is not differentiable in $t$, the large deviation
upper bound holds, but the lower bound may fail (see \cite{Ellis}
for a counterexample).

We now define what we mean by the central limit theorem in our context.

\bd We say that a collection of operators $W_\la$, $\la\subset\Z^d$ satisfies the central limit
theorem if there exists $\si^2 >0$, such that
for all $t\in\R$
\be\label{CLT}
\lim_{\la\uparrow\Z^d}\omega \left(e^{it W_\la}\right) = e^{-t^2 \sigma^2/2}
\ee
\ed

Bryc's theorem~\cite{bryc} gives a connection between the large deviation principle and the central limit theorem.
In our context this means that if $F$ exists in a neighborhood of the origin
in the complex plane, then the central limit theorem (\ref{CLT}) holds,
with  $W_\la = \frac{1}{\sqrt{|\la|}} \sum_{i\in\la} (X_i-\omega (X_i))$  but
possibly $\si^2=0$, in which case the statement is empty. If the sum
\[
\chi^2_X = \sum_{i\in\Z^d} \omega \left(\left(X_i-\omega (X)(X_0-\omega (X)\right)\right)
\]
converges absolutely, then $\si^2= \chi^2_X$.

\subsection{High temperature KMS states}
The states we consider in this paper are KMS states for a translation invariant
finite range potential
at high temperature. This is a collection of self-adjoint $\Phi (A) \in \uu_A$,
indexed by finite
subsets $A\subset\Z^d$ with the following two
properties.
\ben
\item Translation invariance: $\Phi (A +i)= \tau_i \Phi (A)$
\item Finite range: there exists $R>0$ such that if ${\rm diam} (A) > R$, then
$\Phi (A) =0$.
\een
Later on we will see that we can slightly relax the finite range condition, see
(\ref{expot}) below.
The KMS-state associated to the potential $\Phi$ at inverse temperature $\beta$ is
defined as the limit of the finite volume states on $\uu_\la$
\be\label{fvol}
\omega^\beta_\la  (X) = \frac{ \tr (X e^{-\beta H^\Phi_\la})}{  \tr ( e^{-\beta H^\Phi_\la})}
\ee
where the Hamiltonians $H^\Phi_\la$ are defined by
\be
H^\Phi_\la = \sum_{A\subset \la} \Phi (A)
\ee

\br
The KMS-states we consider are defined by the limit
of (\ref{fvol}) as $\la \uparrow\Z^d$. In that way we avoid the
question of uniqueness of KMS-states.

In our context there exists $\beta'_0$ small enough such that there exists a unique
KMS-state, which is possible by the finite range property
(or by its generalization (\ref{expot}), see e.g. proposition 6.2.45 in \cite{Bratrob}, but
this $\beta'_0$ depends on the dimension of the single site algebra, and is possibly smaller
than the $\beta_0$ of our main result stated below.
\er

We can now state our main result.
\bt\label{main}
\ben
\item
There exists $\beta_0$ independent of $X$ such that for all $\beta< \beta_0$
the generating function
\be\label{compgen}
F(z)=
\lim_{\La\uparrow\Z^d} \frac{1}{|\La |} \log \omega
\left( e^{z \sum_{i\in\Lambda} X_i} \right)
\ee
exists and is analytic in a strip $\{ z=x+ iy\in\C: |y| <\delta \}$.
\item
The large deviation principle (\ref{LDP}) holds.
\item
The central limit theorem (\ref{CLT}) holds for the operators
\be
W_\la = \frac{1}{\sqrt{|\la|}}\sum_{i\in\la} (X_i-\omega (X_0)).
\ee
\een
\et

\section{ Non-interacting case: product states}
The simplest situation is the case
\begin{equation}
\omega = \otimes_{i} \omega_i,
\end{equation}
where $\omega_i$ are copies of a faithful state on $M$, i.e., there exists $A$ such that
for $X\in M$:
\begin{equation}
\omega_0 (X) = \frac{ \tr (X e^{-A})}{\tr(e^{-A})}
\end{equation}
The generating function (\ref{ft}) is
\begin{equation}
F(z) = \log \left( \frac{\tr [e^{zX} e^{-A}]}{\tr [e^{-A}]} \right),
\end{equation}
which is clearly defined and analytic
on the strip $\{ z= x+ iy\in\ce: |y| < \delta \}$ for $\delta$ small enough, and
\begin{equation}
\frac{dF}{dz} = \frac{\tr [Xe^{zX} e^{-A}]}{\tr [e^{zX}e^{-A}]}
\end{equation}
In that case the distribution of $X_{\La}/|\La |$ is the same as that of the
$\frac{1}{|\la|}\sum_{i\in\la} \tilde{X_i}$ where $\tilde{X_i}$ are i.i.d.\ with distribution
$\pee_X$. Hence the large deviation principle (\ref{LDP}) is clearly satisfied
with entropy function
\begin{equation}
I(x) = \sup_{t\in \R} (tx- F(t)).
\end{equation}

\section{ Comparison with the classical case}

In the classical Gibbs formalism, there is a natural way to obtain
large deviation probabilities by perturbing the Hamiltonian with a magnetic
field potential (``Cramer tilting").
Let us informally follow this procedure in our context. For simplicity
we put $\beta =1$ in this section.
If we want to know the probability of the event
$\frac{X_{\La}}{|\La |}\simeq a$, then we perturb the hamiltonian $H^\Phi_{\La}$
with an external field $h_a X_{\La}$ to make the value $a$ ``typical", i.e.,
such that
\begin{equation}\label{typ}
\lim_{\La\uparrow\Z^d}\frac{\tr [X_0 e^{- H^\Phi_{\La} - h_a X_{\La}} ]}
{\tr [ e^{- H^\Phi_{\La} - h_a X_{\La}  }]}= a
\end{equation}
The rate function can then (again informally) be obtained as follows:
\begin{eqnarray}\label{info}
&&\omega_{\La} \left( 1_{(a-\epsilon, a+\epsilon )}
(\frac{X_{\La}}{|\La |})\right)
\nonumber\\
&=& \frac{\tr \left( e^{- H^\Phi_{\La} - h_a X_{\La} }
[ e^{H^{\Phi}_{\La} + h_a X_{\La}} e^{- H^\Phi_{\La}}]
1_{(a-\epsilon,a+\epsilon )} ( \frac{X_{\La}}{\La} ) \right)}
{\tr\left( e^{-H^\Phi_{\La}-h_a X_{\La}} \right)}
\frac{\tr [e^{-H^\Phi_{\La} - h_a X_{\La} }]}{\tr [ e^{-H^\Phi_{\La} ]}}.
\end{eqnarray}
Define the pressure
\begin{equation}\label{press}
P( \Phi ) := \lim_{\La \uparrow\Z^d}\frac{1}{|\Lambda|} \log \tr
[e^{-H^\Phi_{\La}} ],
\end{equation}
and
\begin{equation}
P( \Phi, h_a) := \lim_{\Lambda\uparrow\Z^d}
\frac{1}{|\Lambda |}
\log \tr [ e^{-H^\Phi_{\La} - h_a X_{\La}} ].
\end{equation}
We can rewrite (\ref{info}) as
\begin{eqnarray}\label{rewrite}
&&\omega_{\La} \left( 1_{(a-\epsilon, a+\epsilon)}
( \frac{X_{\La}}{|\La |} )\right)
= \frac{\tr [e^{-H^\Phi_{\La} - h_a X_{\La}}
1_{(a-\epsilon,a+\epsilon)} (\frac{X_{\La}}{|\La |} ) ]}
{\tr [e^{-H^\Phi_{\La} - h_a X_{\La}}]} e^{|\Lambda | (a h_a+ P(\Phi, h_a)- P(\Phi))}
e^{o(|\la|)}\nonumber\\
\end{eqnarray}
Since $h_a$ is chosen according to
(\ref{typ}), the first factor in (\ref{rewrite}) is close to one, and
we obtain
\begin{equation}
\log\omega_{\La} \left(1_{(a-\epsilon, a+\epsilon )} (\frac{X_{\La}}{|\La |} )\right)
=
|\La |( a h_a + P(\Phi,h_a)- P(\Phi)) + o(|\la|)
\end{equation}
This suggests as a rate function
\begin{equation}\label{itilde}
\tilde{I} (a) = - [ h_a a + P (\Phi,h_a) - P (\Phi ) ],
\end{equation}
which is the Legendre transform of
\begin{equation}\label{ftilde}
\tilde{F} (t) = \lim_{\La \uparrow\Z^d}\frac{1}{|\La |} \log
\frac{ \tr [e^{t\sum_{i\in\La } X_i - H^\Phi_{\La } }]}
{\tr [e^{-H^\Phi_\Lambda}]}.
\end{equation}
This argument leading to $\tilde{I}, \tilde{F}$ is of course informal,
but in the classical case it is easy and standard to make it rigorous in order to obtain the lower bound.

By the notation $\tilde{I}$ we suggest  that $\tilde{I}$ is not the
entropy function $I$ we are looking for. Indeed, if the large deviation
principle (\ref{LDP}) holds for $X_{\La}/{|\La | }$,
then the only candidate for $I$ is the Legendre transform of
\begin{equation}\label{truef}
F(t) = \lim_{\La\uparrow\Z^d} \frac{1}{|\La |}
\log
\frac{ \tr_{\La}[e^{t\sum_{i\in\La } X_i} e^{-H^\Phi_\Lambda}]}
{\tr_{\La} [e^{-H^\Phi_{\La}}]}
\end{equation}
By the Golden-Thompson inequality we have
\begin{equation}\label{fF}
\tilde{F} (t) \leq F (t),
\end{equation}
and hence
\begin{equation}\label{iI}
\tilde{I} (x) \geq I (x)
\end{equation}
Therefore, the large deviation principle of Theorem \ref{main} implies the following.
\bp\label{entroprop}
For $G\subset \R$ open,
\begin{equation}
\liminf_{\La\uparrow\Z^d} \frac{1}{|\Lambda|} \log \pee^X_{\La} (G)
\geq -\inf_{x\in G} \tilde{I} (x),
\end{equation}
where $\tilde{I} (x)$ is defined in (\ref{itilde}).
\ep

If $X$ and $H^\Phi_\Lambda$ commute, then
(\ref{fF}) becomes an equality and $\tilde{I} (x)$ is actually the true
entropy function, but in the case
$[X, H^{\Phi}_{\La}]\not= 0$, the inequality (\ref{iI}) can be strict.

Notice that even in the simplest case of product states of the previous section,
$I\not= \tilde{I}$ as soon as $A$ and $X$ do not commute.
A possible explanation here is that the ``perturbed states" obtained by adding
a magnetic field potential to the Hamiltonian are not the right states to make the
large deviation event typical.

\section{ Boundary terms}
In the previous section we considered as a candidate generating function
\begin{equation}\label{freef}
F_{f} (t) = \lim_{\La \uparrow\Z^d} \frac{1}{|\La |}
\log
\frac{\tr_{\La} [e^{t\sum_{i\in\Lambda} X_i} e^{-\beta H^{\Phi}_{\La}}]}
{\tr_{\La} [ e^{-\beta H^{\Phi}_{\La}}]},
\end{equation}
where we now add the subindex $f$ to denote free boundary conditions. The
reader might have noticed that we should have written, following
(\ref{ft}):
\begin{eqnarray}\label{echtef}
&&F(t) =
\lim_{\La\uparrow\Z^d} \frac{1}{|\La |} \log \omega
[ e^{t\sum_{i\in\La} X_i} ]\nonumber\\
&=&
\lim_{\La\uparrow\Z^d}\frac{1}{|\La|} \lim_{\La'\uparrow\Z^d} \log
\frac{ \tr_{\La'} \left( \exp \left( t\sum_{i\in\Lambda} X_i\right)
\exp \left (-\beta H^\Phi_{\La'}  \right)\right) }
{\tr_{\La'}\exp \left (-\beta H^\Phi_{\La'}  \right)},
\end{eqnarray}
The difference between $F_f (t)$ and $F (t)$ is caused by a boundary term
and hence it is expected to vanish in the thermodynamic limit,
i.e., we expect that
\begin{equation}\label{hope}
F_f (t) = F (t).
\end{equation}
To be more precise, for $\la'\supset \la$:
\[
H^\Phi_{\La'} = H^\Phi_\la + W^\Phi_{\La,\La'} + H^\Phi_{\La'\setminus\la}
\]
where
\begin{equation}\label{boundary}
W^\Phi_{\La,\La'} = \sum_{A\subset \Lambda', A\cap \Lambda\not=\emptyset,
A\cap\Lambda^c\not=\emptyset} \Phi (A).
\end{equation}
Remark that $\|W^\Phi_{\La,\La'}\| = O (|\partial\la|)$, and since
$H_\La^\Phi$ and $H_{\La'\setminus\la}^\Phi$ commute,
\be
\frac{ \tr_{\La'} \left( \exp \left( t\sum_{i\in\Lambda} X_i\right)
\exp \left (-\beta H^\Phi_\La -\beta H^\Phi_{\La'\setminus\La} \right)\right) }
{\tr_{\La'}\exp \left (-\beta H^\Phi_{\La}-\beta H^\Phi_{\La'\setminus \La } \right)}
=
\frac{\tr_{\La} [e^{t\sum_{i\in\Lambda} X_i} e^{-\beta H^{\Phi}_{\La}}]}
{\tr_{\La} [ e^{-\beta H^{\Phi}_{\La}}]}
\ee
Hence, if we omit the boundary term $W^\Phi_{\La, \La'}$ in (\ref{echtef}), then
we recover $F_f (t)$. The main problem is to omit $W_{\La,\La'}^\Phi$ in the numerator
of (\ref{echtef}), and to prove that the ``price" for this omission is of order $(e^{o(|\la|)})$.

This reminds us on the proof of the existence of the pressure, see e.g. \cite{Israel}, \cite{Simon}.
However, in the quantum case this result relies on the inequality
\begin{equation}
|\log \tr (e^{A+B}) - \log \tr (e^A) | \leq \| B \|.
\end{equation}
In order to prove (\ref{hope}) in a similar way, we would like to have
an estimate like
\begin{equation}\label{logtr}
|\log \omega (e^{A+B} ) - \log \omega (e^A ) |\leq \alpha \| B \|,
\end{equation}
for a state $\omega$, where $\alpha$ does not depend on $A, B$. But such an inequality does not hold!

More precisely, if $ \omega = \tr(e^{H}\cdot)/ \tr (e^H)$ (we omit for a moment the indices $\la$ referring to
the volume), then
\[
|\log \omega (e^{A+B} ) - \log \omega (e^A ) |=\int_{0}^1 \frac{d}{dt} \log \omega (e^{A+tB}) dt
\]
and
\[
\frac{d}{dt} \log \omega (e^{A+tB})
= \frac{\tr \left(\int_0^1 e^{\haa} e^{-s(\haa)}e^H e^{s(\haa)} B\ ds\right)}
{\tr \left(\int_0^1 e^{\haa}e^{-s(\haa)}e^H e^{s(\haa)}\ ds\right)}
=: \Psi (B)
\]
In general $\Psi$ is not a state (unless $A+tB$ and $H$ commute), and the norm of $\Psi$ (as a functional
of $B$) will depend on $A,B$ and $H$, as
the following proposition shows.

\bp
For any $X\in \uu_\la$ with ${\rm Ker }(X)=\{ 0\}$, define
\begin{equation}
\Psi_{X} (B) = \frac{\tr ( X B )}{\tr ( X )}
\end{equation}
$\Psi_X$ defines a continuous functional of $C$ with norm
\begin{equation}
\| \Psi_X \| = \frac{ \tr | X |}{|\tr (X )|} \geq 1,
\end{equation}
with $|X| = \sqrt{ X^* X }$.
In particular, for $X\geq 0$, $\|\Psi_X\|=1$.
\ep
{\bf Proof:} Put $X= J |X|$, where $J$ is a
partial isometry and $|X| = \sqrt{ XX^*} \geq 0 $. Since
${\rm Ker} (X)= \{ 0 \}$, $J$ is
a unitary operator, see \cite{Kad}, Theorem 6.1.2.
Since $|X| \geq 0 $, $\omega_{|X|} (C) : = \tr (\cdot |X| ) / \tr (|X| )$
defines
a state. We have:
\begin{eqnarray}
\left|\frac{\tr (CX ) }{\tr (X)}\right| = \left| \frac{ \omega_{|X|} (CJ )}
{\omega_{|X|} (J)} \right| &\leq & \frac{\| C \| \| J \| }{ | \omega_{|X|} (J ) |}
= \| C \| \left| \frac {\tr [ |X| ]}{\tr [ J|X| ]} \right|\nonumber\\
&=&  \left| \frac{ \tr [|X| ]}{\tr (X) }\right| \| C \|,
\end{eqnarray}
and we obtain
\begin{equation}
\| \Psi_{X} \| \leq \frac{\tr (|X| )}{|\tr ( X )| }.
\end{equation}
If we choose $C= J^*$, then
\begin{equation}
\frac{\tr (CX )}{|\tr X|} =  \frac{\tr (|X|)}{|\tr (X )|},
\end{equation}
so
\begin{equation}
\| \Psi_{X} \| = \frac{\tr ( |X| )}{|\tr (X)|}.
\end{equation}
\QED

This proposition shows that we cannot hope to obtain a useful
version of
(\ref{logtr}) in order to show
(\ref{hope}).  Indeed, if $X$ is not positive
(the $X$ we are thinking about here is $\int_0^1 e^{\haa} e^{-s(\haa)}e^H e^{s(\haa)}$),
then $\|\Psi_X\|$ can be arbitrary large.

Instead we will use a cluster expansion to show the negligibility
of the boundary terms.

\section{Cluster Expansion}

In this section we develop a strategy to prove both existence and analyticity
of (\ref{freef}) and the equality (\ref{hope}), which is based on a quantum cluster expansion. For an introduction to this technique and a comparison of different approaches, see~\cite{Park}. Here we develop a variant of this expansion, by rewriting the partition function of a quantum model as a partition function of a certain (classical) polymer model. Then, the results on the convergence of the expansion follow whenever the Kotecky-Preiss criterion is satisfied~\cite{Kotecky}.

\subsection{Set-up}

Rewrite (\ref{freef})
\begin{eqnarray}
F_f (t) &=&
\lim_{\La\uparrow\Z^d} \frac{1}{|\La |} \log
\frac{\tr_{\La} \left(
e^{t\sum_{i\in\Lambda} X_i} e^{-\beta H^\Phi_\Lambda}\right)}
{\tr_{\La} e^{t\sum_{i\in\La} X_i}}
+ \log \tr (e^{tX}) - P (\beta\Phi )\nonumber\\
&=&
\lim_{\La\uparrow\Z^d} \frac{1}{|\La |}
\log \omega^t_{\La} (e^{-\beta H^\Phi_{\La}} ) + \log \tr (e^{tX} ) - P(\beta\Phi),
\end{eqnarray}
where $\omega^t_{\La}$ is a {\it product} state on $\U_\la$ defined by
\begin{equation}\label{eq: ref-state}
\omega^t_{\La} (Y) =
\frac{\tr_{\Lambda} (e^{t\sum_{i\in\Lambda} X_i} Y)}
{\tr_{\Lambda} (e^{t\sum_{i\in\Lambda}X_i })}.
\end{equation}
The product property of the state $\omega^t_X$ is crucial and due to the fact
that we consider only the averages of a one-point observable. It implies
for $A\in\U_{\La'}$, $B\in\U_{\La''}$, $\La'\cap\La''=\emptyset$:
\begin{equation}
\omega^t_{\Lambda} (A B) = \omega^t_\Lambda (A)\, \omega^t_\Lambda (B).
\end{equation}
This factorization is crucial to set up the cluster expansion that
will allow us to show the existence of the limit
\be
\Xi_f (t)= \lim_{\la\uparrow\Z^d} \frac{1}{|\la|} \log Z^{t,\beta}_\la= \lim_{\la\uparrow\Z^d} \frac{1}{|\la|}\log\omega^t_\la (e^{-\beta H^\Phi_\la})
\ee
Similarly, for $\la'\supset\la$
define
$\omega^t_{\la',\la}$ by
\be
\omega^t_{\la',\la} (Y) = \frac{ \tr_{\la'} (e^{-t\sum_{i\in\la} X_i} Y)}{\tr_{\la'}(e^{-t\sum_{i\in\la} X_i}\otimes
{\rm Id}_{\La'\setminus\la} )}
\ee
which is also a product state, and this time we have
\begin{eqnarray}
  F(t) &=& \lim_{\La\uparrow\Z^d}  \frac{1}{|\La|} \lim_{\La'\uparrow\Z^d} \log
  \frac{\tr_{\La'} \bigl( e^{t\sum_{i\in\La} X_i}\,e^{-\beta H^\Phi_{\La'}} \bigr)}
  {\tr_{\La'} \bigl( e^{t\sum_{i\in\La} X_i} \otimes \text{Id}_{\La' \setminus \La}
  \bigr)}
  \frac{\tr_{\La} \bigl( e^{t\sum_{i\in\La} X_i} \bigr)}
  {\tr_{\La} \bigl(e^{-\beta H^\Phi_\La} \bigr)}
  \frac{\tr_{\La'} \bigl(
  e^{-\beta H^\Phi_\La} \otimes \text{Id}_{\La' \setminus \La} \bigr)}
  {\tr_{\La'} \bigl( e^{-\beta H^\Phi_{\La'}} \bigr)}
  \nonumber
\\
  &=& \log \tr(e^{tX}) - P(\beta\Phi)
  + \lim_{\La\uparrow\Z^d} \frac{1}{|\La|} \lim_{\La' \uparrow \Z^d}
  \log \frac{\om_{\La',\La}^t(e^{-\beta H^\Phi_{\La'}})\,
  \om_{\La}^0 (e^{-\beta H^\Phi_\La})}
  {\om_{\La'}^0 (e^{-\beta H^\Phi_{\La'}})}
\end{eqnarray}
where $\om_{\La'}^0 = \om_{\La'}^{t=0}$ is the trace state on $\U_{\La'}$.
The existence of $F(t)$ is equivalent with the existence of
\begin{equation}
  \Xi (t) = \lim_{\la\uparrow\Z^d} \frac{1}{|\la|} \lim_{\La'\uparrow\Z^d}
  \log \tilde{Z}^{t,\beta}_{\La',\La}
\end{equation}
where
\begin{equation}\label{eq: tilde-Z}
  \tilde Z^{t,\beta}_{\La',\La} =
  \frac{\om_{\La',\La}^t(e^{-\beta H^\Phi_{\La'}})\,
  \om_{\La}^0 (e^{-\beta H^\Phi_\La})}
  {\om_{\La'}^0 (e^{-\beta H^\Phi_{\La'}})}
\end{equation}
Moreover,  the equality $F_f(t) = F(t) $ will follow from $\Xi_f (t) = \Xi (t)$.

Our strategy is then described as follows.
\ben
\item Set up the cluster expansion in order to define $\Xi(t)$, $\Xi_f (t)$.
This can be done by properly defining a polymer model
and by using the Kotecky-Preiss criterion.
\item Equality of $\Xi$ and $\Xi_f$ follows from the fact that in the
expansion, only clusters touching the boundary of $\la$ will make
the difference between $\log Z^{t,\beta}_\la$ and
$\log \tilde{Z}^{t,\beta}_{\La',\La}$
\item Analyticity is proved by showing that the polymer weights
are analytic in $t$ and satisfy the Kotecky-Preiss criterion in
a strip in the complex plane.
\een

\subsection{Polymer model}

In order to compute $\Xi_f(t)$, we use the idea of the Mayer expansion and rewrite the finite volume expectation $\om_\La^t(e^{-\beta H_\La^\Phi})$ as the partition function of a  polymer gas. Due to the product structure of the state, the polymer weights become  independent up to the exclusion, and we can use familiar results on the convergence of a series for the logarithm of such partition functions.

We start by writing the series
\begin{equation}\label{eq: exp}
  Z_\La^{t,\beta} = \om_\La^t \Bigl[ \sum_{n=0}^{\infty}
  \frac{(-\beta H_\La)^n}{n!} \Bigr]
  = \om_\La^t \Bigl[ \sum_{n=0}^{\infty} \frac{(-\beta)^n}{n!}
  \sum_{A_1,\ldots,A_n \subset \La} \Phi(A_1) \ldots \Phi(A_n) \Bigr]
\end{equation}
that can be cast into the form of a polymer expansion as follows. We use the notation $\Ga = (A_1,\ldots,A_n)$ for any finite sequence of finite sets of sites and the shorthand $\Phi_\Ga = \Phi(A_1)\ldots\Phi(A_n)$. Let $G_\Ga$ be the graph over the set of vertices $\{1,\ldots,n\}$ such that $1\leq i < j \leq n$ are connected by edge whenever $A_i \cap A_j \neq \emptyset$. A sequence
$\Ga' = (A'_1,\ldots,A'_k)$ is called a maximally connected subsequence of $\Ga$ whenever
there is a maximally connected component of the graph $G_\Ga$ with the vertex set
$i_1 < \ldots < i_k$ such that $A'_1 = A_{i_1},\ldots,A'_k = A_{i_k}$.
The following lemma is then an immediate application of these definitions.
\begin{lemma}
Let
$\{\ga_\al\}_{\al \in I}$ be the collection of all maximally connected subsequences of $\Ga$. Then $\Phi_{\Ga} = \prod_{\al \in I} \Phi_{\ga_{\al}}$ and the product does not depend on the order.
\end{lemma}
Connected sequences of sets are called polymers and we use the symbol $\caA_\La$ for
the set of all polymers in $\La$.
Any sequence (respectively set) of polymers
$(\ga_1,\ldots,\ga_n)$, $\ga_\al = (A^\al_{k_1},\ldots,A^\al_{k_\al})$,
$\al = 1,\ldots,n$
is called an admissible sequence (respectively set) if
$A^\al_i \cap A^{\al'}_j = \emptyset$ for any $i,j$ and $\al \neq \al'$. Given any sequence of sets $\Ga$, the collection
$\{\ga_\al\}_{\al\in I}$ of all maximally connected components of $\Ga$
is clearly an admissible set of polymers, but the correspondence is obviously not one-to-one. A simple observation is that there are exactly
\[
  \frac{(\sum_\al k(\al))!}{\prod_\al k(\al)!}
\]
sequences $\Ga$ such that $\{\ga_\al\}_{\al\in I}$ is the collection of all maximally connected subsequences of $\Ga$.
Defining $|\Ga| = k$ for any sequence $\Ga = (A_1,\ldots,A_k)$, we rewrite $Z_\La^{t,\beta}$ as the partition function of a polymer model:
\begin{eqnarray}
  Z_\La^{t,\beta} &=& \sum_{n=0}^\infty \frac{(-\beta)^n}{n!}
  \sum_{\Ga:\,|\Ga| = n} \prod_\al \om_\La^t(\Phi_{\ga_\al})
  \nonumber
\\
  &=& \sum_{n=0}^{\infty}
  \sum_{k=0}^{\infty} \frac{1}{k!} \sum_{l_1,\ldots,l_k \geq 1 \atop \sum_i l_i = n}
  \sum_{\ga_1,\ldots,\ga_k \in \caA_\La \atop |\ga_1| = l_1,\ldots,|\ga_k| = l_k}
  g(\ga_1,\ldots,\ga_n)
  \prod_{i=1}^{k} \Bigl[ \frac{(-\beta)^{l_i}}{l_i!} \om_\La^t(\Phi_{\ga_i}) \Bigr]
  \nonumber
\\
  &=& \sum_{n=0}^{\infty} \frac{1}{n!} \sum_{\ga_1,\ldots,\ga_n \in \caA_\La}
  g(\ga_1,\ldots,\ga_n)\,\prod_{i=1}^{n} \rho^{t,\beta}(\ga_i)
\end{eqnarray}
where we have introduced the weights
\begin{equation}
  \rho^{t,\beta}(A_1,\ldots,A_k)
  = \frac{(-\beta)^k}{k!} g_C(A_1,\ldots,A_k)\,
  \om_\La^t(\Phi(A_1)\ldots\Phi(A_k))
\end{equation}
and the indicator functions
\begin{eqnarray}
  g(\ga_1,\ldots,\ga_n) &=&
  \begin{cases}
    1  &  \text{if $(\ga_1,\ldots,\ga_n)$ is admissible}
  \\
    0  &  \text{otherwise}
  \end{cases}
\\
  g_C(A_1,\ldots,A_k) &=&
  \begin{cases}
    1  &  \text{if $(A_1,\ldots,A_k)$ is connected}
  \\
    0  &  \text{otherwise}
  \end{cases}
\end{eqnarray}
Note that the polymers have been defined as sequences of sets rather than collections of sets. Obviously, the weight $\rho^{t,\beta}(\ga = A_1,\ldots,A_k)$ generically depends on the order of the sets $A_1,\ldots,A_k$, whereas it does not depend on $\La$ as far as
$\La \supset \cup_{i=1}^{k} A_i$, due to the product structure of the state
$\om_\La^t$.
The cluster expansion now reads \cite{Kotecky,Miracle-Sole}, formally,
\begin{eqnarray}
  \log Z_\La^{t,\beta} &=&
  \sum_{n=1}^{\infty} \frac{1}{n!} \sum_{\ga_1,\ldots,\ga_n \in\caA_\La}
  a_T(\ga_1,\ldots,\ga_n) \prod_{i=1}^{n} \rho^{t,\beta}(\ga_i)
  \nonumber
\\
  &=& \sum_{C \subset \La}  \sum_{n=1}^{\infty} \frac{1}{n!}
  \sum_{\ga_1,\ldots,\ga_n \in\caA_C \atop \text{Supp}(\ga_1,\ldots,\ga_n) = C}
  a_T(\ga_1,\ldots,\ga_n) \prod_{i=1}^{n} \rho^{t,\beta}(\ga_i)
  \nonumber
\\
\label{eq: cluster-exp}
  &=& \sum_{C \subset \La} w^{t,\beta}(C)
\end{eqnarray}
where we have defined the ``cluster'' weights $w^{t,\beta}(C)$ by the partial resummation over all sequences
$(\ga_1,\ldots,\ga_n)$, $\ga_\al = (A^\al_1,\ldots,A^\al_{k(\al)})$, $n \geq 1$, such that
$\text{Supp}(\ga_1,\ldots,\ga_n) := \cup_{\al=1}^n \cup_{i=1}^{k(\al)} A^\al_i = C$.

Next we set up a similar expansion for $\tilde Z_{\La',\La}^{t,\beta}$.
Note first that by taking $t = 0$ in \eqref{eq: cluster-exp} we immediately get cluster expansions for both expectations $\om^0_\La(e^{-\beta H^\Phi_\La})$ and
$\om^0_{\La'}(e^{-\beta H^\Phi_{\La'}})$. For
$\om_{\La',\La}^{t}(e^{-\beta H^\Phi_{\La'}})$ we repeat the same steps to get
\begin{equation}
  \om_{\La',\La}^t(e^{-\beta H^\Phi_{\La'}}) =
  \sum_{n=0}^{\infty} \frac{1}{n!} \sum_{\ga_1,\ldots,\ga_n \in \caA_{\La'}}
  g(\ga_1,\ldots,\ga_n)\,\prod_{i=1}^{n} \tilde \rho^{t,\beta}_\Lambda(\ga_i)
\end{equation}
with the notation
\begin{equation}
  \tilde \rho^{t,\beta}_\La(A_1,\ldots,A_k)
  = \frac{(-\beta)^k}{k!} g_C(A_1,\ldots,A_k)\,
  \om_{\La',\La}^t(\Phi(A_1)\ldots\Phi(A_k))
\end{equation}
In particular, $\tilde\rho_\La^{t,\beta}(A_1,\ldots,A_k)$ does not depend on $\La'$ if
$\cup_i A_i \subset \La'$, and
\begin{equation}\label{eq: coincidence}
  \tilde\rho_\La^{t,\beta}(A_1,\ldots,A_k) =
  \begin{cases}
     \rho^{t,\beta}(A_1,\ldots,A_k)  &  \text{if } \cup_i A_i \subset \La
   \\
     \rho^{0,\beta}(A_1,\ldots,A_k)  &  \text{if } \cup_i A_i \subset
     \La' \setminus \La
  \end{cases}
\end{equation}
Defining
\begin{equation}
  \tilde w_\La^{t,\beta}(C) =
  \sum_{n=1}^{\infty} \frac{1}{n!}
  \sum_{\ga_1,\ldots,\ga_n \atop \text{Supp}(\ga_1,\ldots,\ga_n) = C}
  a_T(\ga_1,\ldots,\ga_n)\,
  \prod_{i=1}^{n} \tilde\rho_\La^{t,\beta}(\ga_i)
\end{equation}
we also have $\tilde w_\La^{t,\beta}(C) = w^{t,\beta}(C)$ for any $C \subset \La$ and
$\tilde w_\La^{t,\beta}(C) = w^{0,\beta}(C)$ for any $C \subset \La' \setminus \La$.
Hence, we get from \eqref{eq: tilde-Z} that
\begin{eqnarray}\label{eq: cluster-exp-tilde}
  \log Z_{\La',\La}^{t,\beta} &=&
  \sum_{C\subset\La'} \tilde w_\La^{t,\beta}(C)
  + \sum_{C\subset\La} w^{0,\beta}(C)
  - \sum_{C\subset\La'} w^{0,\beta}(C)
  \nonumber
\\
  &=& \sum_{C \subset \La} w^{t,\beta}(C)
  + \sum_{C \subset \La' \atop C \not\subset \La,\, C \not\subset \La' \setminus \La}
  \tilde w_{\La}^{t,\beta}(C)
\end{eqnarray}
where the first term coincides with the series for $\log Z_\La^{t,\beta}$ and the second
one is a boundary term summing up the clusters intersecting both sets $\La$ and
$\La' \setminus \La$.

\vspace{10mm}
The existence of the cluster weights $w^{t,\beta}(C)$ and $w_\La^{t,\beta}(C)$ and upper bounds can be proven
under the assumption that the polymer weights are sufficiently damped. We show the Kotecky-Preiss criterion of the convergence \cite{Kotecky,Miracle-Sole} to be satisfied in a high-temperature regime:
\begin{lemma}\label{lem: geometric}
Let $a,\beta_0 > 0$ such that
\begin{equation}\label{eq: potential-cond}
  \sum_{B \ni 0} e^{2a|B|}\, (e^{\beta_0\|\Phi_B\|} - 1) \leq a
\end{equation}
Then one has the upper bounds
\begin{equation}
  \sup_{(t,\beta) \in \R \times [0,\beta_0]}
  \sum_{n=1}^{\infty} \sum_{A_1,\ldots,A_n \atop 0 \in \cup_i A_i}
  e^{a|\cup_i A_i|} |\rho^{t,\beta}(A_1,\ldots,A_n)| \leq a
\end{equation}
and
\begin{equation}
  \sup_\La \sup_{(t,\beta) \in \R \times [0,\beta_0]}
  \sum_{n=1}^{\infty} \sum_{A_1,\ldots,A_n \atop 0 \in \cup_i A_i}
  e^{a|\cup_i A_i|} |\tilde\rho_\La^{t,\beta}(A_1,\ldots,A_n)| \leq a
\end{equation}
\end{lemma}
Combining with Theorem~2 in \cite{Miracle-Sole}, we have the following result.
\begin{proposition}\label{prop: cluster-bound}
Under Condition~\eqref{eq: potential-cond}, the cluster weights $w^{t,\beta}(C)$ are
translation-invariant and
\begin{equation}
  \sup_{(t,\beta) \in \R \times [0,\beta_0]}
  \sum_{C \ni 0} |w^{t,\beta}(C)| \leq a
\end{equation}
Similarly,
\begin{equation}
  \sup_\La \sup_{(t,\beta) \in \R \times [0,\beta_0]}
  \sum_{C\ni 0} |\tilde w_{\La}^{t,\beta}(C)| \leq a
\end{equation}
\end{proposition}
\begin{proof}[Proof of Lemma~\ref{lem: geometric}]
Using the notation $\un\ga$ for the image of the sequence
$\ga = (A_1,\ldots,A_k)$, we have the upper bound
\begin{eqnarray}
  \sum_{\ga:\, \un\ga = \{B_1,\ldots,B_l\}} |\rho^{t,\beta}(\ga)| &\leq&
  g_C(B_1,\ldots,B_l)\, \sum_{n=1}^{\infty} \frac{\beta^n}{n!}
  \sum_{k_1,\ldots,k_l \geq 1 \atop k_1 + \ldots + k_l = n}
  {n \choose k_1,\ldots,k_l} \prod_{i=1}^{l} \|\Phi(B_i)\|^{k_i}
  \nonumber
\\
  &=& g_C(B_1,\ldots,B_l)\, \prod_{i=1}^{l} \sum_{k=1}^{\infty}
  \frac{\beta^k}{k!} \|\Phi(B_i)\|^k
  \nonumber
\\
  \label{eq: geom-1}
  &=& g_C(B_1,\ldots,B_l)\, \prod_{i=1}^{l} (e^{\beta\|\Phi(B_i)\|} - 1)
\end{eqnarray}
for any collection of finite sets $\{B_1,\ldots,B_l\}$, $l=1,2,\ldots$ The identical
upper bound holds if we replace $\rho^{t,\beta}$ with $\tilde\rho_\La^{t,\beta}$.
Hence, we only need to concentrate on the former weight, for which we have the inequality
\begin{equation}\label{eq: collection}
  \sum_{n=1}^{\infty} \sum_{A_1,\ldots,A_n \atop 0 \in \cup_i A_i}
  e^{a |\cup_i A_i|} |\rho^{t,\beta}(A_1,\ldots,A_n)|
  \leq \sum_{l=1}^\infty \sum_{\{B_1,\ldots,B_l\} \atop 0 \in \cup_i B_i}
  g_C(B_1,\ldots,B_l) \prod_{i=1}^{l} \phi(B_i)
\end{equation}
with the shortcut
\begin{equation}
  \phi(B) = e^{a|B|}\,(e^{\beta\|\Phi(B)\|} - 1)
\end{equation}
In order to estimate the right-hand side of~\eqref{eq: collection}, we consider the function
\begin{eqnarray}
  Y(\Delta \,|\,B) &=& \sum_{l\geq 1}
  \sum_{B \in \{B_1,\ldots,B_l\} \subset \Delta}
  g_C(B_1,\ldots,B_l) \prod_{i=1}^{l} \phi(B_i)
\\
  &=& \phi(B) \sum_{l\geq 0} \sum_{\{B_1,\ldots,B_l\} \subset \Delta \setminus \{B\}}
  g_C(B,B_1,\ldots,B_l) \prod_{i=1}^{l} \phi(B_i)
\end{eqnarray}
defined on all pairs of a finite set $B$ and a finite collection $\Delta$ of finite sets, $\Delta \ni B$. For this function we derive a recurrent inequality, realizing that
the collection $\{B_1,\ldots,B_l\}$ obtained from a connected collection
$\{B,B_1,\ldots,B_l\}$ splits in general into a number of connected components, each of them intersecting $B$. By upper-bounding the sums over these components separately we subsequently write:
\begin{eqnarray}
  Y(\Delta \,|\, B) &\leq& \phi(B) \sum_{m\geq 0} \frac{1}{m!}
  \sum_{l_1,\ldots,l_m \geq 1} \prod_{\al=1}^{m} \Bigl(
  \sum_{\{B_1,\ldots,B_{l_\al}\} \subset \Delta \setminus \{B\}
  \atop B \cap \cup_i B_i \neq \emptyset}
  g_C(B_1,\ldots,B_{l_\al}) \prod_{i=1}^{l_\al} \phi(B_i) \Bigr)
  \nonumber
\\
  &\leq& \phi(B) \sum_{m\geq 0} \frac{1}{m!} \Bigl(
  \sum_{D:\, D \cap B \neq \emptyset} \phi(D) \sum_{l \geq 0}
  \sum_{\{B_1,\ldots,B_l\} \subset \Delta \setminus \{B,D\}}
  g_C(D,B_1,\ldots,B_l) \Bigr)^m
  \nonumber
\\
 &\leq& \phi(B) \exp \Bigl[ |B| \sup_x \sum_{D \ni x \atop D \in \Delta \setminus \{B\}}
 Y(\Delta \setminus \{B\} \,|\, D) \Bigr]
 \label{eq: recurrence}
\end{eqnarray}
Since
\begin{equation}
  \sum_{n=1}^{\infty} \sum_{A_1,\ldots,A_n \atop 0 \in \cup_i A_i}
  e^{a |\cup_i A_i|} |\rho^{t,\beta}(A_1,\ldots,A_n)|
  \leq \sup_\Delta \sum_{B \ni 0 \atop B \in \Delta} Y(\Delta \,|\, B)
\end{equation}
the proof of the lemma is finished by proving by induction in the cardinality of $\Delta$ that
\begin{equation}
  \sup_x \sum_{B \ni x \atop B \in \Delta} Y(\Delta \,|\, B) \leq a
\end{equation}
uniformly in $\Delta$.
Indeed, by the induction hypothesis and using inequalities~\eqref{eq: potential-cond}, \eqref{eq: recurrence} and the translation-invariance of the potential, we get
\begin{eqnarray}
  \sup_x \sum_{B \ni x \atop B \in \Delta} Y(\Delta \,|\, B)  &\leq&
  \sup_x \sum_{B \ni x} \phi(B)\, e^{a|B|}
  = \sum_{B \ni 0} e^{2a|B|}\,(e^{\beta\|\Phi(B)\|} - 1) \leq a
\end{eqnarray}
The case $\Delta = \{\emptyset\}$ is trivial.
\end{proof}

\subsection{Analyticity of the polymer weights}

Under a slightly stronger condition than \eqref{eq: potential-cond},
we prove the existence of an analytic continuation of the polymer weights
$\rho^{z,\beta}(\ga)$ to a set
$\{z = x + iy \in \C;\,|y| < \delta\} \times \{\beta;\,|\beta| < \beta_0\}$, uniformly for all polymers. Since the linear functional $\om_\La^{z}$ given by
formula~\eqref{eq: ref-state} is not a state anymore for $z \not\in \R$ due to the lack of positivity, we write $\rho^{z,\beta}(\ga)$ in the form
\begin{eqnarray}
  \rho^{z=x+iy,\beta}(A_1,\ldots,A_k) &=&
  \frac{(-\beta)^k}{k!} g_C(A_1,\ldots,A_k)\,\om_\La^{z}
  (\Phi(A_1)\ldots\Phi(A_k))
  \nonumber
\\
  &=& \frac{(-\beta)^k}{k!} g_C(A_1,\ldots,A_k)
\\
  &\phantom{-}& \times
  \Bigl(\frac{\tr e^{x X}}{\tr e^{(x + iy)X}} \Bigr)^{|\La|}
  \om_\La^{x}(\Phi(A_1)\ldots\Phi(A_k)\,e^{i\delta\sum_{i\in\La} X_i}),
  \nonumber
\end{eqnarray}
an identity valid for any $\La \supset \cup_{i=1}^{k} A_i$. Choosing $\La = \cup_{i=1}^k A_i$ and using the bound
\begin{eqnarray}
  \frac{|\tr\, e^{(x+iy)X}|}{\tr\, e^{x X}} &=&
  |\om^{x}_{\{0\}}(e^{iy X})|
  = |\om^{x}_{\{0\}}(\sum_{n=0}^{\infty}\frac{(iy)^n}{n!}X^n)|
\\
  &\geq& 1 - \sum_{n=1}^{\infty} \frac{|y|^n}{n!} \|X\|^n =
  2 - e^{|y|\|X\|}
\end{eqnarray}
we obtain the next variant of inequality~\eqref{eq: geom-1}:
\begin{eqnarray}
  \sum_{\ga:\, \un\ga = \{B_1,\ldots,B_l\}} |\rho^{x + iy,\beta}(\ga)|
  &\leq& g_C(B_1,\ldots,B_l)\, \Bigl( \frac{1}{2-e^{|y| \|X\|}}
  \Bigr)^{|\cup_{i=1}^{l} B_l|}
  \prod_{i=1}^{l} (e^{|\beta|\|\Phi_{B_i}\|} - 1)
  \nonumber
\\
  &\leq& g_C(B_1,\ldots,B_l)\, \prod_{i=1}^{l}
  \frac{e^{|\beta|\|\Phi_{B_i}\|} - 1}{(2-e^{|y|\|X\|})^{|B_i|}}
\end{eqnarray}
The other steps in the proof of Lemma~\ref{lem: geometric} remain unchanged if we replace there the function $\phi(B)$ with
\begin{equation}
  \phi'(B) = \Bigl( \frac{e^{a}}{2-e^{|y|\|X\|}} \Bigr)^{|B|}\,
  (e^{|\beta|\|\Phi(B)\|} - 1)
\end{equation}
and assume condition~\eqref{eq: potential-cond-an} below. As a result, we get the upper bound on the analytic continuation of the polymer weights as
\begin{equation}
  \sup_{|y| < \delta,\,|\beta| < \beta_0}
  \sum_{n=1}^{\infty} \sum_{A_1,\ldots,A_n \atop 0 \in \cup_i A_i}
  e^{a|\cup_i A_i|} |\rho^{z=x+iy,\beta}(A_1,\ldots,A_n)| \leq a
\end{equation}
The uniform bounds on the cluster weights in the region $|y| < \delta_0$,
$|\beta| < \beta_0$ then follow by \cite{Kotecky,Miracle-Sole}. As a consequence, the   (partially resummed cluster) weights $w^{z,\beta}(C)$ are analytic by Vitali's theorem and we arrive at the following result:
\begin{proposition}\label{prop: analyticity}
Assume there are $a,\delta,\beta_0 > 0$ such that
\begin{equation}\label{eq: potential-cond-an}
  \sum_{B \ni 0}
  \Bigl( \frac{e^{2a}}{2 - e^{\delta\|X\|}}\Bigr)^{|B|}\,
  (e^{\beta_0\|\Phi_B\|} - 1) \leq a
\end{equation}
Then the cluster weights $w^{z,\beta}(C)$ are analytic in the region
\[
  \caV_{\delta,\beta_0} = \{(z,\beta)\in\C^2:\,|\text{Im}\,z| < \delta,\,
  |\beta| < \beta_0\}
\]
for all finite sets of sites $C$. Moreover,
\begin{equation}\label{eq: complex}
  \sup_{(z,\beta) \in \caV_{\delta,\beta_0}}
  \sum_{C \ni 0} |w^{z,\beta}(C)| \leq a
\end{equation}
\end{proposition}

\section{Proof of Theorem \ref{main}}

One easily checks that under the assumption
\be\label{expot}
\sum_{B \ni 0} e^{\epsilon |B|}\,\|\Phi(B)\| < \infty
\ee
for some $\epsilon > 0$, there exist $a,\delta,\beta_0 > 0$ such that
condition~\eqref{eq: potential-cond-an} is satisfied. Moreover, $\beta_0$
can be chosen independent of $X$.

{\bf Existence of $F_f(t)$}.\\
It follows from the cluster expansion for $\log Z_\La^{t,\beta}$ due to the translation-invariance of the cluster weights $w^{t,\beta}(C)$. To see this, write
\begin{equation}
  \log Z_\La^{t,\beta} - |\La| \sum_{C \ni 0} \frac{w^{t,\beta}(C)}{|C|}
  = -\sum_{i\in\La} \sum_{C \ni i \atop C \not\subset \La} \frac{w^{t,\beta}(C)}{|C|}
\end{equation}
By Proposition~\ref{prop: cluster-bound}, $\sum_{C\ni 0} |w^{t,\beta}(C)| \leq a$, and
for any $\epsilon > 0$ there exists a finite set of sites $\caD$ such that
$\sum_{C \ni 0,\,C \not\subset \caD} |w^{t,\beta}(C)| \leq \epsilon$.
Introducing
\begin{equation}
  \La_0 = \{i\in\La;\,\caD + i \subset \La\}
\end{equation}
we have
\begin{eqnarray}
  \bigl| \log Z_\La^{t,\beta} - |\La| \sum_{C \ni 0} \frac{w^{t,\beta}(C)}{|C|} \bigr|
  &\leq&  \bigl( \sum_{i\in\La_0} + \sum_{i \in \La \setminus \La_0} \bigl)
  \sum_{C \ni i \atop C \not\subset \La} |w^{t,\beta}(C)|
\\
  &\leq& \epsilon|\La_0| + a|\La \setminus \La_0|
\end{eqnarray}
and the limit $\lim_{\epsilon\downarrow 0} \lim_{\La \uparrow \Z^d}$, the latter being taken in the van Hove sense so that $\lim |\La_0|/|\La| = 1$, yields
\begin{equation}\label{eq: final}
  \Xi_f(t) = \lim_{\La\uparrow\Z^d} \frac{1}{|\La|} \log Z_\La^{t,\beta}
  = \sum_{C \ni 0} \frac{w^{t,\beta}(C)}{|C|}
\end{equation}
Moreover, we have got the upper bound $\sup_{t\in\R} |\Xi_f(t)| \leq a$.

{\bf Equality $F_f(t) = f(t)$.}\\
Notice first that the limit $\La'\uparrow\Z^d$ exists,
\begin{equation}
  \lim_{\La'\uparrow\Z^d} \log Z_{\La',\La}^{t,\beta} =
  \log Z_{\La}^{t,\beta} + \sum_{C \not\subset \La,\, C \cap \La \neq \emptyset}
  \tilde w_\La^{t,\beta}(C)
\end{equation}
by the absolute convergence of the second sum. Using the same argument as above,
the second term is of order $o(|\La|)$, and we get the equality $\Xi(t) = \Xi_f(t)$.

{\bf Analyticity of $F_f(t)$}.\\
We only need to prove the analyticity of the function $\Xi_f(t)$ given by
the series~\eqref{eq: final}. By Proposition~\ref{prop: analyticity}, all cluster weights
have an analytic continuation to the strip $|\text{Im}\,z| < \delta_0$. Since the series converges there uniformly due to \eqref{eq: complex}, $\Xi_f(z)$ is analytic in the strip by Vitali's theorem.

Finally, the existence and the differentiability of $F(t)$ implies both large deviation upper and lower bounds by Gartner-Ellis theorem. Since $F(t)$ has an analytic continuation to a neighborhood of the origin, Bryc's theorem implies the central limit theorem. To see that $\si^2 >0$ for $\beta$ small enough, consider first
the case $\beta=0$, then
\[
\si^2 = \left(\frac{d^2}{dt^2} F(t)\right)_{t=0} = \omega_0(X^2)- \omega_0 (X)^2
\]
where $\omega_0$ is the normalized trace.
Hence in that case, $\si^2>0$ as soon as $X$ has non-trivial spectrum. Therefore, by the
convergence of the cluster expansion, the variance  $\si^2_\beta = \si_0^2 +  O(\beta)$
is strictly positive for $\beta$ small enough. Moreover, it is given by the absolutely converging
sum
\[
\si_\beta^2 = \sum_{i\in\Z^d} \omega\left( (X_i -\omega (X_i))(X_0-\omega (X_0))\right)
\]


\subsection{A generalization }

Our result on the convergence of the cluster expansions for $\Xi_f(t)$ and $\Xi(t)$ can be slightly generalized.
We sketch this generalization here without too much details.  Let $\{\Phi_k\}_{k=1,\ldots,n}$ be a family of potentials and $\om$ be a product state. Then the generating function
\begin{equation}
  Z_\La^z = \om(e^{z_1 H^{\Phi_1}_\La} \ldots e^{z_n H^{\Phi_n}_\La})
\end{equation}
where $z= (z_1,\ldots,z_n) \in \C^n$, admits a cluster expansion
\begin{equation}
  \log Z_\La^{z} = \sum_{C \subset \La} w^z(C)
\end{equation}
with the cluster weights $w^z(C)$ depending only on 
$\Phi_i(A), A \subset C, i = 1,\ldots,n$,
and one has the following result.
\begin{proposition}\label{genprop}
Assume that
\begin{equation}\label{eq: general}
  \sup_x \sum_{B \ni x} e^{2a|B|} (e^{\sum_{i=1}^{n} \delta_i \|\Phi_i(B)\|} - 1) \leq a
\end{equation}
for some $a,\delta_1,\ldots,\delta_n > 0$. Then all cluster weights
$w^z(C)$ are analytic in the polydisc
$\caD = \{z = (z_1,\ldots,z_n):\,|z_i| \leq \delta_i,\,i=1,\ldots,n\}$ and
\begin{equation}
  \sup_{z \in \caD} \sup_x \sum_{C \ni x} |w^z(C)| \leq a
\end{equation}
\end{proposition}
\begin{remark}
\begin{enumerate}
\item
Notice that \eqref{eq: general} is the same condition one would write for the convergence of the cluster expansion for $\log \om(\exp\sum_i z_i H^{\Phi_i}_\la)$.
\item
As a corollary, one obtains the existence and analyticity in a neighborhood of the origin for various kinds of (cumulant) generating functions. In particular, by taking $\Phi_1 = \Phi$ translationally invariant, $z_1 = \beta$, $z_2 = z$, and
$\Phi_2(B) = \sum_{i\in\Z^d} \tau_i(X)\, 1_{B = D + i}$ for a fixed set of sites $D$ and an operator $X \in \caU_{D}$, one gets the existence and analyticity for the (free b.c.) generating function
\[
  F_f^X(z) = \lim_{\La\uparrow\Z^d} \frac{1}{|\La|} \log \om_\La^{\Phi,\beta}
  (e^{z\sum_{i:\, D+i \subset \La} \tau_i(X)})
\]
where
\[
  \om_\La^{\Phi,\beta}(Y) = \frac{\tr(e^{-\beta H_\La^{\Phi}}\,Y)}
  {\tr(e^{-\beta H_\La^{\Phi}})}
\]
\item
A necessary and sufficient condition on the potentials $\Phi_1,\ldots,\Phi_n$ to satisfy  \eqref{eq: general} with some $a,\delta_1,\ldots,\delta_n > 0$ is
that there exists $\epsilon > 0$ such that
\[
\sup_x \sum_{B \ni x} e^{\epsilon |B|} \|\Phi_i(B)\| < \infty
\qquad i = 1,\ldots,n
\]
\end{enumerate}
\end{remark}
Proposition \ref{genprop} does not give the (full) large deviation principle since the modulus of
$z_i$ has to be small. However, it does give the central limit theorem for
\[
\frac{H^\Phi_\la- \omega (H^\Phi_\la)}{\sqrt{|\la|}}
\]
because for that we only need analyticity in a neighborhood of $0$.

\section{ Level two large deviations}
We will now define a random measure which can thought of as the distribution
under the state $\omega$ of the ``measure"
$\frac{1}{|\la|} \sum_{i\in\la} \delta_{X_i}$. For $f\in\ce ([-\|X\|, \|X\|],\R)$, and
$\mu$ a probability measure on $[-\|X\|, \|X\|]$, we write
$\langle \mu, f \rangle = \int f d\mu$.

More precisely, for $f_1,\ldots, f_k$ a finite collection of continuous
functions on $[-\|X\|, \|X\|]$, and $A_1,\ldots A_k$ Borel sets, define
\be
\pee ( \langle\loc_\la, f_1\rangle\in A_1, \ldots, \langle\loc_\la, f_k\rangle\in A_k)
= \omega\Bigl( 1_{A_1} \Bigl(\frac{1}{|\la|}\sum_{i\in\la}f_1 (X_i)\Bigr)
\ldots 1_{A_k} \Bigl(\frac{1}{|\la|}\sum_{i\in\la}f_k (X_i)\Bigr)\Bigr)
\ee
This formula defines the distribution of a random measure $\loc_\la$. Indeed,
the sets
\be
\{ \langle\loc_\la, f_1\rangle\in A_1, \ldots \langle\loc_\la, f_k\rangle\in A_k: f_i\in \ce ([-\|X\|,\|X\|],\R), A_i\in \bee \}
\ee
are Borel sets in the weak topology on $\mee_1 (\shit)$, the set of probability measures on $[-\|X\|, \|X \|]$,
and they are generating for the Borel-$\si$-field on $\mee_1 (\shit)$

The candidate level-2 generating function is then a functional on $\ce (\shit, \R)$, given by
\be\label{2gen}
\Psi (f) =\lim_{\la\uparrow\Z^d} \frac1{|\la|}\log\E \left( e^{\langle\loc_\la, f\rangle}\right)
=\lim_{\la\uparrow\Z^d} \frac{1}{|\la|}\log\omega \left( e^{\sum_{i\in\la}f(X_i)}\right)
\ee
And the corresponding candidate large deviation entropy function is its Legendre
transform:
\be\label{2ent}
\iii(\mu ) = \sup_{f\in\ce(\shit,\R)} (\langle\mu,f\rangle - \Psi (f)
\ee
We then have the following theorem

\bt
Suppose $\omega$ is a high temperature KMS state as in Theorem \ref{main}.
\ben
\item
The limit defining the generating function $\Psi (f)$ in (\ref{2gen}) exists
and defines a convex $\Psi: \ce (\shit, \R) \to \R$.
\item The random measures $\loc_\la$ satisfy the large deviation principle
with rate function $\iii$ given by (\ref{2ent}).
\item The relation between $\iii$ and $I$ is given by the contraction principle:
\be
I (x) = \inf \{ \iii (\mu) : \int_{\shit} \xi \mu (d\xi) = x \}
\ee
\een
\et
\bpr
The existence of the limit defining $\Psi$ follows from Theorem \ref{main} and
the fact that $\beta_0$ does not depend on $X$, so we can replace
$X_i$ by $f(X_i)$.

The large deviation principle follows from G\^{a}teaux
differentiability of $\Psi$. More precisely, for any
$f,g\in\ce (\shit,\R)$, the limit
\be\label{gat}
\partial_g \Psi (f) = \lim_{t \to 0}\frac{\Psi (f+ t g)- \Psi (f)}{t}
\ee
exists.
This can be seen as follows. By the same argument showing the analyticity of
$F(z)$ (of \ref{compgen}) in a strip $ \{ z= x+ iy: |y|< \delta\}$ one sees that
$z\mapsto \Psi (f + zg)$ exists and is analytic in a strip
$ \{ z= x+ iy: |y|< \delta\}$, where now $\delta$ depends on $f$ and $g$. This
is clearly sufficient to have the existence of
the limit (\ref{gat}). Then we can apply
Corollary 4.5.27 of \cite{Dembo} to conclude the large deviation principle.

Finally, the contraction principle follows from the fact that the distribution of
$X_\la/|\la|$ is the distribution of $\int \xi \loc_\la (d\xi)$, hence we are in the situation
of the standard contraction principle, \cite{Dembo}, Theorem 4.2.1.
\epr

\section*{Acknowledgements}

We are very grateful to A.~C.~D.~van Enter and C.~Maes for fruitful discussions and many useful comments.


\begin{thebibliography}{99}
\bibitem{bryc}
W.~Bryc,
\newblock A remark on the connection between the large deviation principle
and the central limit theorem, 
\emph{Stat.~and Prob.~Lett.} {\bf 18}:253--256 (1993).

\bibitem{Bratrob}
O.~Bratteli and D.~W.~Robinson,
{\em Operator Algebras and Quantum Statistical Mechanics 2} 
(Springer-Verlag, Berlin, 1996).

\bibitem{Dembo} 
A.~Dembo and O.~Zeitouni, 
\newblock {\em Large Deviations Techniques and Applications} 
\newblock (Springer-Verlag, New York, 1998).

\bibitem{efs}
A.~C.~D.~van Enter, R.~Fern\'{a}ndez, and A.~D.~Sokal, 
\newblock Regularity properties and pathologies of position-space renormalization 
group transformations: Scope and limitations of Gibbsian theory,
\newblock \emph{J.~Stat.~Phys.} {\bf 72}:879--1167 (1993).

\bibitem{geo}
H.-O.~Georgii.
\newblock {\em Gibbs Measures and Phase Transitions}
\newblock (Walter de Gruyter \& Co., Berlin, 1988).

\bibitem{Ellis} 
R.~S.~Ellis, 
\newblock {\em Large Deviations and Statistical Mechanics}
\newblock (Springer-Verlag, New York, 1985).

\bibitem{Israel} 
R.~B.~Israel, 
\newblock {\em Convexity in the Theory of Lattice Gases}
(Princeton University Press, 1979).

\bibitem{Kad} 
R.~V.~Kadison, J.~R.~Ringrose, 
\newblock {\em Fundamentals of the Theory of Operator Algebras} 
\newblock (Academic Press, New York, London 1983).

\bibitem{Kotecky}
\newblock R.~Kotecky and D.~Preiss,
\newblock Cluster {expansion} for {abstract} {polymer} {models},
\newblock {\em Commun.~Math.~Phys.} {\bf103}:491--498 (1986).

\bibitem{Lebowitz} 
J.~L.~Lebowitz, M.~Lenci, and H.~Spohn, 
\newblock Large deviations for ideal quantum systems, 
\newblock (math-phys archive 9906014).

\bibitem{lebspo}  G.~Gallavotti, J.~L.~Lebowitz, and V.~Mastropietro, 
\newblock Large deviations in rarefied quantum gases, 
\newblock \emph{J.~Stat.~Phys.} {\bf 108}:831--861 (2002).

\bibitem{Miracle-Sole}
S.~Miracle-Sol\'e,
\newblock On the {convergence} of {cluster} {expansions},
\newblock {\em Physica A} {\bf 279}:244--249 (2000).

\bibitem{olla}
S.~Olla,
\newblock Large deviations for Gibbs random fields,
\newblock \emph{Prob.~Th.~Rel.~Fields} {\bf 77}:343--357 (1988).

\bibitem{Park} 
Y.~M.~Park, 
\newblock The cluster expansion for classical and quantum lattice systems, 
\emph{J.~Stat.~Phys.} {\bf 27}:553--576 (1982).

\bibitem{Simon} 
B.~Simon, 
\newblock{\em The Statistical Mechanics of Lattice Gases}
(Princeton University Press, 1993).

\bibitem{vets} 
D.~Goderis and P.~Vets,
\newblock Central limit theorem for mixing quantum systems and the CCR-algebra of fluctuations, 
\newblock \emph{Comm.~Math.~Phys.} {\bf 122}:249--265 (1989).

\bibitem{vetsv} 
D.~Goderis, A.~Verbeure, and P.~Vets, 
\newblock Noncommutative central limits,
\newblock \emph{Prob.~Th.~Rel.~Fields} {\bf 82}:527--544 (1989).


\end{thebibliography}
\end{document}